\documentclass{iopart}
\usepackage{iopams}
\usepackage{euscript,amssymb,epsfig}

\usepackage{amsfonts,latexsym}
\usepackage{euscript,amssymb}
\usepackage{epsfig}

\usepackage{graphicx}

\newcommand{\be}[1]{\begin{equation}\label{#1}}
\newcommand{\ee}{\end{equation}}
\newcommand{\ba}[1]{\begin{eqnarray}\label{#1}}
\newcommand{\ea}{\end{eqnarray}}
\newcommand{\rf}[1]{(\ref{#1})}
\newcommand{\nn}{\nonumber}

\begin{document}

\title{Classical tests of multidimensional gravity: negative result}

\author{Maxim Eingorn and Alexander Zhuk}

\address{Astronomical Observatory and Department of
Theoretical Physics, Odessa National University, Street Dvoryanskaya 2, Odessa 65082, Ukraine}

\eads{\mailto{maxim.eingorn@gmail.com} and \mailto{ai$\_$zhuk2@rambler.ru}}


\begin{abstract} In Kaluza-Klein model with toroidal extra dimensions, we obtain the metric coefficients in a weak field approximation for delta-shaped matter sources. These metric coefficients are applied to calculate the formulas for frequency shift, perihelion shift, deflection of light and parameterized post-Newtonian (PPN) parameters. In the leading order of approximation, the formula for frequency shift coincides with well known general relativity expression. However, for perihelion shift, light deflection and PPN parameter $\gamma $ we
obtain formulas $D\pi r_g/[(D-2)a(1-e^2)]$,  $(D-1)r_g/[(D-2)\rho]$ and $1/(D-2)$ respectively, where $D$ is a total number of spatial dimensions. These expressions demonstrate good
agreement with experimental data only in the case of ordinary three-dimensional $(D=3)$ space. This result does not depend on the size of the extra dimensions. Therefore, in considered multidimensional Kaluza-Klein models the point-like masses cannot produce gravitational field
which corresponds to the classical gravitational tests.
\end{abstract}


\maketitle


\section{\label{sec:1}Introduction}

\setcounter{equation}{0}

The idea of the multidimensionality of our Universe demanded by the theories of unification of the fundamental interactions is one of the most breathtaking ideas of
theoretical physics. It takes its origin from the pioneering papers by Th.Kaluza and O.Klein \cite{KK} and now
the most self-consistent modern theories of unification such as superstrings, supergravity and M-theory are constructed
in spacetime with extra dimensions \cite{Polchinski}. Different aspects of the idea of the multidimensionality are intensively used in numerous
modern articles.
Therefore, it is very important to suggest experiments which can reveal the extra dimensions. For example, one of the aims of Large Hadronic Collider consists in
detecting of Kaluza-Klein particles which correspond to excitations of the internal spaces (see e.g. \cite{KKparticles}).
On the other hand, if we can show that the existence of the extra dimensions is contrary to observations, then these theories are
prohibited.
This important problem is extensively discussed in recent scientific literature (see e.g. \cite{KWE}-\cite{Poplawski}).

It is well known that classical gravitational tests such as frequency shift, perihelion shift, deflection of light and time delay of radar echoes
(the Shapiro time delay effect) are crucial tests of any gravitational theory. For
example, there is the significant discrepancy for Mercury between the measurement value of the perihelion shift and its calculated value using Newton's formalism
\cite{Shapiro}. It indicates that non-relativistic Newton's theory of gravity is not complete. This problem was resolved with the help of general relativity which
is in good agreement with observations. Similar situation happened with deflection of light \cite{light deflection}. The Shapiro time delay effect is used to get an
upper limit for the parameterized post-Newtonian parameter $\gamma$ \cite{Bertotti}. Obviously, multidimensional gravitational theories should also be in concordance
with these experimental data. To check it, the corresponding estimates were carried out in a number of papers. For example, in \cite{indians}, it was investigated the
well known multidimensional black hole solution \cite{MP} and the authors obtained a negative result. However, this result was clear from the very beginning because the
solution \cite{MP} does not have non-relativistic Newtonian limit in the case of extra dimensions. Definitely, in solar system such solutions lead to results which are
far from the experimental data. The 5-D soliton metrics \cite{Kramer}-\cite{Davidson} were explored in \cite{KWE}-\cite{LOW}. In papers \cite{limOvWesson} and \cite{LO},
it was found  the range of parameters for which classical gravitational tests for these metrics satisfy the observational values. The black string (see e.g. \cite{GL})
is a particular limiting case of
such solutions with a trivial metric coefficient for the extra dimension. However, it can be easily shown that such solutions do not correspond to the point-like matter
sources.

In 5-D non-factorizable brane world model, classical gravitational tests were investigated in \cite{brane}. Here, the model contains one free parameter associated with
the bulk Weyl tensor. For appropriate values of this parameter, the perihelion shift in this model does not contradict observations. Certainly, this result is of
interest and it is necessary to examine carefully this model to verify the naturalness of the conditions imposed.

In our paper we consider classical gravitational tests in Kaluza-Klein models (factorizable geometry) with an arbitrary number of spatial dimensions $D\ge 3$. We suppose
that in the absence of gravitating  masses the metric is a flat one. Gravitating point-like masses (moving or at rest) perturb this metric and we consider these
perturbations in a weak field approximation. In this approximation, we obtain the asymptotic form of the metric coefficients.  Then we admit that, first, the extra
dimensions are compact and have the topology of tori and, second, gravitational potential far away from gravitating masses tends to non-relativistic Newtonian limit.
In the case of a gravitating mass at rest, the obtained metric coefficients are used to calculate frequency shift, perihelion shift, deflection of light and
parameterized post-Newtonian (PPN) parameters. We demonstrate that for the frequency shift type experiment it is hardly possible to observe the difference between the
usual four-dimensional general relativity and multidimensional Kaluza-Klein models. However, the situation is quite different for perihelion shift, deflection of light
and PPN parameters. In these cases we get formulas which generalize the corresponding ones in general relativity. We show that formulas for perihelion shift, deflection
of light and PPN parameter $\gamma$ depend on a total number of spatial dimensions and they are
in good agreement with observations only in ordinary three-dimensional space. It is important to note that this
result does not depend explicitly on the size of the extra dimensions\footnote{In the leading order of approximation, our formulas do not depend on
sizes of the extra dimensions. All correction terms, where the sizes of the extra dimensions appear, are exponentially suppressed.}.
So, we cannot avoid the problem with classical gravitational tests in a limit of arbitrary small (but non-zero!) size of the extra dimensions. It is worth noting that in paper
\cite{XuMa} the authors arrived at the same conclusions in spite of they use the different approach.

Therefore, our results show that in considered multidimensional Kaluza-Klein models the point-like gravitating masses cannot produce gravitational field
which corresponds to the classical gravitational tests.

The paper is structured as follows. In section 2 we get the asymptotic metric coefficients in the weak field limit for the delta-shaped matter gravitating source. These
metric coefficients are applied to calculate the formulas of frequency shift, perihelion shift, deflection of light and PPN parameters in section 3. The main results are summarized in the
concluding section 4.


\section{\label{sec:2}Weak gravitational field approximation}

\setcounter{equation}{0}

To start with, we consider the general form of the multidimensional metric:
\be{1} ds^2=g_{ik}dx^idx^k=g_{00}\left(dx^0\right)^2+2g_{0\alpha}dx^0dx^{\alpha}+g_{\alpha\beta}dx^{\alpha}dx^{\beta}\, , \ee
where the Latin indices $i,k = 0,1,\ldots ,D$ and the Greek indices  $\alpha ,\beta = 1,\ldots ,D$. $D$ is the total number of spatial dimensions.
We make the natural assumption that in the case of the absence of matter sources the spacetime is Minkowski spacetime: $g_{00}=\eta_{00}=1$,
$g_{0\alpha}=\eta_{0\alpha}=0$, $g_{\alpha\beta}=\eta_{\alpha\beta}=-\delta_{\alpha\beta}$. At the same time, the extra dimensions
may have the topology of tori. In the presence of matter, the metric is not a Minkowskian one and we will investigate it in the weak field limit. This means that the
gravitational field is weak and the velocities of the test bodies are small compared to the speed of light $c$. In this case the metric is only slightly perturbed from
its flat spacetime value:
\be{2} g_{ik}\approx\eta_{ik}+h_{ik}\, , \ee
where $h_{ik}$ are corrections of the order $1/c^2$. In particular, $h_{00} \equiv 2\varphi /c^2$. Later we will demonstrate that $\varphi $ is the non-relativistic
gravitational potential. The same conclusion with respect to $\varphi$ can be easily obtained from the comparison of the non-relativistic action of a test mass moving in
a gravitational field with its relativistic action.
To get the other correction terms up to the same order $1/c^2$, we should consider the multidimensional Einstein equation
\be{3} R_{ik}=\frac{2S_D\tilde G_{\mathcal{D}}}{c^4}\left(T_{ik}-\frac{1}{D-1}g_{ik}T\right)\, , \ee
where $S_D=2\pi^{D/2}/\Gamma (D/2)$ is the total solid angle (surface area of the $(D-1)$-dimensional sphere of unit radius), $\tilde G_{\mathcal{D}}$ is the
gravitational constant in the $(\mathcal{D}=D+1)$-dimensional spacetime. We are going to investigate the weak field approximation where gravitational field is generated
by $N$ moving point masses. Therefore, the energy-momentum tensor is
\be{4} T^{ik}=\sum\limits_{p=1}^Nm_p\left[(-1)^Dg\right]^{-1/2}\frac{dx^i}{dt}\frac{dx^k}{dt}\frac{cdt}{ds}\delta({\bf r}-{\bf r}_{p})\, , \ee
where $m_p$ is the rest mass and ${\bf r}_p$ is the radius vector of the $p$-th particle respectively. All radius vectors ${\bf r}$ and ${\bf r}_{p}$ are
$D$-dimensional, e.g. ${\bf r} =(x^1,x^2,\ldots ,x^D)$ where $x^{\alpha}$ are coordinates in metric \rf{1}. The rest mass density is
\be{5} \rho\equiv \sum\limits_{p=1}^Nm_p\, \delta({\bf r}-{\bf r}_p)\, . \ee

\subsection{$1/c^2$ correction terms}

Obviously, to hold in the right hand side of \rf{3} the terms up to the order $1/c^2$, the components of energy-momentum tensor \rf{4} are approximated as
\be{6} T_{00}\approx\rho c^2,\ \ \ T_{0\alpha}\approx0,\ \ \ T_{\alpha\beta}\approx0 \quad \Rightarrow \quad T=T^i_i\approx \rho c^2\, . \ee

Taking into account that $h_{ik}$ are of the order of $1/c^2$, the covariant components of the Riemann and Ricci tensors
\ba{7} R_{iklm}&=&\frac{1}{2}\left(\frac{\partial^2g_{im}}{\partial x^k\partial x^l}+\frac{\partial^2g_{kl}}{\partial x^i\partial x^m}-\frac{\partial^2g_{il}}{\partial
x^k\partial x^m}-\frac{\partial^2g_{km}}{\partial x^i\partial x^l}\right)\nn \\&+&g_{np}\left(\Gamma^n_{kl}\Gamma^p_{im}-\Gamma^n_{km}\Gamma^p_{il}\right),\ \ \
R_{km}=g^{il}R_{iklm}\ea
up to the same order read correspondingly:
\be{8} R_{iklm}\approx\frac{1}{2}\left(\frac{\partial^2h_{im}}{\partial x^k\partial x^l}+\frac{\partial^2h_{kl}}{\partial x^i\partial
x^m}-\frac{\partial^2h_{il}}{\partial x^k\partial x^m}-\frac{\partial^2h_{km}}{\partial x^i\partial x^l}\right)\, ,\ee
\ba{9} R_{km}&\approx&\frac{1}{2}\eta^{il}\left(\frac{\partial^2h_{im}}{\partial x^k\partial x^l}+\frac{\partial^2h_{kl}}{\partial x^i\partial
x^m}-\frac{\partial^2h_{il}}{\partial x^k\partial x^m}-\frac{\partial^2h_{km}}{\partial x^i\partial x^l}\right)\nn \\
&=&\frac{1}{2}\left(\frac{\partial^2h_m^l}{\partial x^k\partial x^l}+\frac{\partial^2h_k^l}{\partial x^m\partial x^l}-\frac{\partial^2h_l^l}{\partial x^k\partial
x^m}-\eta^{il}\frac{\partial^2h_{km}}{\partial x^i\partial x^l}\right)\, , \ea
where $h^i_k\equiv\eta^{im}h_{mk}$. With the help of the gauge conditions
\be{10} \frac{\partial}{\partial x^k}\left(h_i^k-\frac{1}{2}h_l^l\delta_i^k\right)=0\, , \ee
the formula \rf{9} can be written in the form
\be{11} R_{km}\approx-\frac{1}{2}\eta^{il}\frac{\partial^2h_{km}}{\partial x^i\partial x^l}\, . \ee
Taking into account that the derivatives with respect to $x^0=ct$ are much smaller than the derivatives with respect to $x^{\alpha}$, we obtain from \rf{11}:
\be{12}R_{00}\approx-\frac{1}{2}\eta^{\alpha\beta}\frac{\partial^2h_{00}}{\partial x^{\alpha}\partial
x^{\beta}}=\frac{1}{2}\delta^{\alpha\beta}\frac{\partial^2h_{00}}{\partial x^{\alpha}\partial x^{\beta}}=\frac{1}{2}\triangle h_{00},\ee\
\be{13}R_{0\alpha}\approx\frac{1}{2}\triangle h_{0\alpha},\ \ \ R_{\alpha\beta}\approx\frac{1}{2}\triangle h_{\alpha\beta}\,\ee
where $\triangle = \delta^{\alpha\beta}\partial^2/\partial x^{\alpha}\partial x^{\beta}$ is the $D$-dimensional Laplace operator. It is worth noting that for the
condition \rf{10} up to the order $1/c^2$ holds
\be{14} \frac{\partial}{\partial x^{\beta}}\left(h_{\alpha}^{\beta}-\frac{1}{2}h_l^l\delta_{\alpha}^{\beta}\right)=0 +O(1/c^3)\, ,\ \quad \frac{\partial
h_0^{\beta}}{\partial x^{\beta}}=0+O(1/c^3)\, . \ee

Therefore, keeping in the left hand and right hand sides of \rf{3} terms up to the order $1/c^2$ we obtain the following equations:
\ba{15}
\triangle h_{00}&=&\frac{2S_DG_{\mathcal{D}}}{c^2}\rho\, ,\quad \triangle h_{0\alpha}=0\, , \nn \\
\triangle h_{\alpha\beta}&=&\frac{1}{D-2}\cdot\frac{2S_DG_{\mathcal{D}}}{c^2}\rho\delta_{\alpha\beta}\, , \ea
where $G_{\mathcal{D}}=[2(D-2)/(D-1)]\, \tilde G_{\mathcal{D}}$. Substitution of $h_{00} = 2\varphi /c^2$ into the above equation for $h_{00}$ demonstrates that $\varphi
$ satisfies the $D$-dimensional Poisson equation:
\be{16} \triangle \varphi = S_DG_{\mathcal{D}}\rho\, . \ee
Therefore, $\varphi$ is the non-relativistic gravitational potential. From \rf{15} we obtain
\be{17} h_{0\alpha}=0\, ,\quad h_{\alpha\beta}=\frac{1}{D-2}\, h_{00}\delta_{\alpha\beta}=\frac{1}{D-2}\, \frac{2\varphi}{c^2}\delta_{\alpha\beta}\, . \ee
It can be easily seen that in this approximation spacial coordinates of the metric \rf{1} are the isotropic ones, i.e. the spacial part of the metric is conformally
related to the Euclidean one.  It is worth noting also that the relation $h_{\alpha\beta}/h_{00}=[1/(D-2)]\delta_{\alpha\beta}$ can be also obtained from the
corresponding equations in papers \cite{MP,CDandMP}.

\subsection{$1/c^3$ and $1/c^4$ correction terms}

Now, we want to keep in metric \rf{1} the terms up to the order $1/c^2$. Because the coordinate $x^0=ct$ contains $c$, this means that in $g_{00}$ and $g_{0\alpha}$ we
should keep correction terms up to the order $1/c^4$ and $1/c^3$ respectively and to leave $g_{\alpha\beta}$ without changes in the form $g_{\alpha\beta}
\approx\eta_{\alpha\beta}+h_{\alpha\beta}$ with $h_{\alpha\beta}$ from \rf{17}.

First, we investigate the energy-momentum tensor \rf{4} which we split into three expressions:
\be{18} T^{00}=\sum\limits_{p=1}^Nm_pc^2[(-1)^Dg]^{-1/2}\frac{cdt}{ds}\delta({\bf r}-{\bf r}_p)\, , \ee
\be{19} T^{0\alpha}=\sum\limits_{p=1}^Nm_pc[(-1)^Dg]^{-1/2}v_p^{\alpha}\frac{cdt}{ds}\delta({\bf r}-{\bf r}_p)\, , \ee
\be{20} T^{\alpha\beta}=\sum\limits_{p=1}^Nm_p[(-1)^Dg]^{-1/2}v_p^{\alpha}v_p^{\beta}\frac{cdt}{ds}\delta({\bf r}-{\bf r}_p)\, ,\ee
where $v_p^{\alpha}=dx_p^{\alpha}/dt\, $. From \rf{20} we obtain up to order 1 (in units $c$) the covariant components
\be{21} T_{\alpha\beta}\approx\sum\limits_{p=1}^Nm_pv_{p\alpha}v_{p\beta}\delta({\bf r}-{\bf r}_p)\, . \ee
Thus, taking into account the prefactor $1/c^4$ in the right hand side of \rf{3}, these components can contribute to $g_{\alpha\beta}$ terms of the order of $1/c^4$
which is not of interest for us. For $T_{0\alpha}$ we find from \rf{19}:
\be{22} T_{0\alpha}\approx-\sum\limits_{p=1}^Nm_pcv_{p\alpha}\delta({\bf r}-{\bf r}_p)\, .\ee
Hence, these components can give in $g_{0\alpha}$ terms  of the order of $1/c^3$ which is of interest for us. Finally, for $T_{00}$ we get from \rf{18}:
\ba{23} T_{00}=g_{0i}g_{0k}T^{ik}\approx(g_{00})^2T^{00}\nn \\ \fl\approx\left(1+\frac{2\varphi}{c^2}\right)^2\left[(-1)^D\left(1+\frac{2\varphi}{c^2}\right)
\left(-1+\frac{1}{D-2}\frac{2\varphi}{c^2}\right)^D\right]^{-1/2}\nn \\
\fl\times\sum\limits_{p=1}^Nm_pc^2\left[\left(1+\frac{2\varphi}{c^2}\right)-\frac{v_p^2}{c^2}\right]^{-1/2}\delta({\bf r}-{\bf r}_p)\approx
\left(1+\frac{4\varphi}{c^2}\right)\left(1+\frac{1}{D-2}\frac{2\varphi}{c^2}\right)\nn \\
\fl\times\sum\limits_{p=1}^Nm_pc^2\left(1-\frac{\varphi}{c^2}+\frac{v_p^2}{2c^2}\right)\delta({\bf r}-{\bf
r}_p)\approx\sum\limits_{p=1}^Nm_pc^2\left(1+\frac{3D-4}{D-2}\frac{\varphi}{c^2}+\frac{v_p^2}{2c^2}\right)\delta({\bf r}-{\bf r}_p)\nn \\
\fl= \sum\limits_{p=1}^Nm_pc^2\delta({\bf r}-{\bf r}_p)+\sum\limits_{p=1}^Nm_p\left(\frac{3D-4}{D-2}\varphi_p+\frac{1}{2}v_p^2\right)\delta({\bf r}-{\bf r}_p)\, , \ea
%
where $\varphi_p$ is potential of gravitational field in a point with radius vector ${\bf r}_p$. At the moment, we do not care about the fact that $\varphi_p$ contains
the infinite contribution of the $p$-th particle. Thus, up to order 1 we get
\ba{24} T&=&g^{ik}T_{ik}\approx g^{00}T_{00}+g^{\alpha\beta}T_{\alpha\beta}\approx\sum\limits_{p=1}^Nm_pc^2\delta({\bf r}-{\bf r}_p)\nn \\
&+&\sum\limits_{p=1}^Nm_p\left(\frac{D}{D-2}\varphi_p-\frac{1}{2}v_p^2\right)\delta({\bf r}-{\bf r}_p)\, . \ea
With the help of \rf{23} and \rf{24} we obtain up to the order $1/c^4$:
\ba{25} &{}&\frac{2S_D\tilde G_{\mathcal{D}}}{c^4}\left(T_{00}-\frac{1}{D-1}g_{00}T\right)\approx
\frac{S_DG_{\mathcal{D}}}{c^2}\sum\limits_{p=1}^Nm_p\delta({\bf r}-{\bf r}_p)\nn \\
&+&\frac{S_DG_{\mathcal{D}}}{c^4}\sum\limits_{p=1}^Nm_p\left(\frac{3D-4}{D-2}\varphi_p+\frac{D}{2(D-2)}v_p^2\right)\delta({\bf r}-{\bf r}_p). \ea
Similarly, from \rf{22} and \rf{24}  we get up to the order $1/c^3$:
\be{26} \fl\frac{2S_D\tilde G_{\mathcal{D}}}{c^4}\left(T_{0\alpha}-\frac{1}{D-1}g_{0\alpha}T\right)\approx
-\frac{D-1}{D-2}\frac{S_DG_{\mathcal{D}}}{c^3}\sum\limits_{p=1}^Nm_pv_{p\alpha}\delta({\bf r}-{\bf r}_p)\, . \ee

Now, we shall work out the left hand side of \rf{3} up to an appropriate orders of $1/c$. As we wrote above, we are looking for corrections of the order of $1/c^4$ and
$1/c^3$ to the metric components $g_{00}$ and $g_{0\alpha}$, respectively. To this end, it is convenient to present $g_{ik}$ as follows:
\be{27} g_{ik}\approx\eta_{ik}+h_{ik}+f_{ik}\, , \ee
where $f_{00}$ and $f_{0\alpha}$ are of the order of $1/c^4$ and $1/c^3$, respectively. Then, the Riemann curvature tensor \rf{7} reads
\ba{28} R_{iklm}\approx\frac{1}{2}\left(\frac{\partial^2h_{im}}{\partial x^k\partial x^l}+\frac{\partial^2h_{kl}}{\partial x^i\partial
x^m}-\frac{\partial^2h_{il}}{\partial x^k\partial x^m}-\frac{\partial^2h_{km}}{\partial x^i\partial x^l}\right)\nn \\
\fl+\frac{1}{2}\left(\frac{\partial^2f_{im}}{\partial x^k\partial x^l}+\frac{\partial^2f_{kl}}{\partial x^i\partial x^m}-\frac{\partial^2f_{il}}{\partial x^k\partial
x^m}-\frac{\partial^2f_{km}}{\partial x^i\partial x^l}\right)+\eta^{np}\left(\Gamma_{n,kl}\Gamma_{p,im}-\Gamma_{n,km}\Gamma_{p,il}\right)\nn \\
\fl\approx\frac{1}{2}\left(\frac{\partial^2h_{im}}{\partial x^k\partial x^l}+\frac{\partial^2h_{kl}}{\partial x^i\partial x^m}-\frac{\partial^2h_{il}}{\partial
x^k\partial x^m}-\frac{\partial^2h_{km}}{\partial x^i\partial x^l}\right)\nn \\
\fl+\frac{1}{2}\left(\frac{\partial^2f_{im}}{\partial x^k\partial x^l}+\frac{\partial^2f_{kl}}{\partial x^i\partial x^m}-\frac{\partial^2f_{il}}{\partial x^k\partial x^m}-\frac{\partial^2f_{km}}{\partial x^i\partial x^l}\right)\nn \\
\fl+\frac{1}{4}\eta^{np}\left(\frac{\partial h_{nk}}{\partial x^l}+\frac{\partial h_{nl}}{\partial x^k}-\frac{\partial h_{kl}}{\partial x^n}\right)\left(\frac{\partial
h_{pi}}{\partial x^m}+\frac{\partial h_{pm}}{\partial x^i}-\frac{\partial h_{im}}{\partial x^p}\right)\nn \\
\fl-\frac{1}{4}\eta^{np}\left(\frac{\partial h_{nk}}{\partial x^m}+\frac{\partial h_{nm}}{\partial x^k}-\frac{\partial h_{km}}{\partial x^n}\right)\left(\frac{\partial
h_{pi}}{\partial x^l}+\frac{\partial h_{pl}}{\partial x^i}-\frac{\partial h_{il}}{\partial x^p}\right)\, . \ea
From this formula we obtain the Ricci tensor:
\ba{29} R_{km}\approx-\frac{1}{2}\eta^{il}\frac{\partial^2h_{km}}{\partial x^i\partial x^l}+\frac{1}{2}H_{km}-\frac{1}{2}\eta^{il}\frac{\partial^2f_{km}}{\partial
x^i\partial x^l}+\frac{1}{2}F_{km}\nn \\
-\frac{1}{2}\eta^{ij}\eta^{lp}h_{jp}\left(\frac{\partial^2h_{im}}{\partial x^k\partial x^l}+\frac{\partial^2h_{kl}}{\partial x^i\partial
x^m}-\frac{\partial^2h_{il}}{\partial x^k\partial x^m}-\frac{\partial^2h_{km}}{\partial x^i\partial x^l}\right)\nn \\
+\frac{1}{4}\eta^{il}\eta^{np}\left(\frac{\partial h_{nk}}{\partial x^l}+\frac{\partial h_{nl}}{\partial x^k}-\frac{\partial h_{kl}}{\partial
x^n}\right)\left(\frac{\partial h_{pi}}{\partial x^m}+\frac{\partial h_{pm}}{\partial x^i}-\frac{\partial h_{im}}{\partial x^p}\right)\nn \\
-\frac{1}{4}\eta^{il}\eta^{np}\left(\frac{\partial h_{nk}}{\partial x^m}+\frac{\partial h_{nm}}{\partial x^k}-\frac{\partial h_{km}}{\partial
x^n}\right)\left(\frac{\partial h_{pi}}{\partial x^l}+\frac{\partial h_{pl}}{\partial x^i}-\frac{\partial h_{il}}{\partial x^p}\right)\, , \ea
where we introduced the notations:
\ba{30} H_{km}&=&\eta^{il}\left(\frac{\partial^2h_{im}}{\partial x^k\partial x^l}+\frac{\partial^2h_{kl}}{\partial x^i\partial x^m}-\frac{\partial^2h_{il}}{\partial
x^k\partial x^m}\right)\nn \\
&=&\frac{\partial^2h_m^0}{\partial x^k\partial x^0}+\frac{\partial^2h_k^0}{\partial x^0\partial x^m}+\frac{\partial^2h_m^{\beta}}{\partial x^k\partial
x^{\beta}}+\frac{\partial^2h_k^{\beta}}{\partial x^{\beta}\partial x^m}-\frac{\partial^2h_l^l}{\partial x^k\partial x^m} \ea
and
\ba{31} F_{km}&=&\eta^{il}\left(\frac{\partial^2f_{im}}{\partial x^k\partial x^l}+\frac{\partial^2f_{kl}}{\partial x^i\partial x^m}-\frac{\partial^2f_{il}}{\partial
x^k\partial x^m}\right)\nn \\
&=&\frac{\partial^2f_m^0}{\partial x^k\partial x^0}+\frac{\partial^2f_k^0}{\partial x^0\partial x^m}+\frac{\partial^2f_m^{\beta}}{\partial x^k\partial
x^{\beta}}+\frac{\partial^2f_k^{\beta}}{\partial x^{\beta}\partial x^m}-\frac{\partial^2f_l^l}{\partial x^k\partial x^m}\, . \ea
Taking into account that $h^{\alpha}_0=h_{\alpha}^0\equiv 0$, we get
\be{32} H_{00}=\frac{\partial^2h_0^0}{\partial(x^0)^2}-\frac{\partial^2h_{\alpha}^{\alpha}}{\partial(x^0)^2} \ee
and
\be{33} H_{0\alpha}=\frac{\partial^2h_{\alpha}^{\beta}}{\partial x^0\partial x^{\beta}}-\frac{\partial^2h_{\beta}^{\beta}}{\partial x^0\partial x^{\alpha}}\, , \ee
which are of the order $1/c^4$ and $1/c^3$, respectively. For the components $F_{km}$ we obtain
\be{34} F_{00}\approx2\frac{\partial^2f_0^{\beta}}{\partial x^0\partial x^{\beta}}=\frac{\partial^2h_{\beta}^{\beta}}{\partial(x^0)^2}\, ,\quad
F_{0\alpha}\approx\frac{\partial^2f_0^{\beta}}{\partial x^{\beta}\partial x^{\alpha}}=\frac{1}{2}\frac{\partial^2h_{\beta}^{\beta}}{\partial x^0\partial x^{\alpha}}\, ,
\ee
which are defined up to the orders  $1/c^4$ and $1/c^3$, respectively. To get these expressions, we use the gauge condition
\be{35} \frac{\partial f_0^{\beta}}{\partial x^{\beta}}=\frac{1}{2}\frac{\partial h_{\beta}^{\beta}}{\partial x^0}\, . \ee
Therefore, the 00-component of the Ricci tensor reads
\ba{36} R_{00}&\approx&\frac{1}{2}\triangle h_{00}
-\frac{1}{2}\frac{\partial^2h_{00}}{\partial(x^0)^2}+\frac{1}{2}\frac{\partial^2h_0^0}{\partial(x^0)^2}-\frac{1}{2}\frac{\partial^2h_{\alpha}^{\alpha}}{\partial(x^0)^2}\nn
\\
&+&\frac{1}{2}\triangle f_{00}+\frac{1}{2}\frac{\partial^2h_{\beta}^{\beta}}{\partial(x^0)^2}+\frac{1}{2}\eta^{ij}\eta^{lp}h_{jp}\frac{\partial^2h_{00}}{\partial x^i\partial x^l}\nn \\
&+&\frac{1}{4}\eta^{il}\eta^{np}\left(\frac{\partial h_{n0}}{\partial x^l}-\frac{\partial h_{0l}}{\partial x^n}\right)\left(\frac{\partial h_{p0}}{\partial
x^i}-\frac{\partial h_{i0}}{\partial x^p}\right)\nn \\
&-&\frac{1}{4}\eta^{il}\eta^{np}\left(-\frac{\partial h_{00}}{\partial x^n}\right)\left(\frac{\partial h_{pi}}{\partial x^l}+\frac{\partial h_{pl}}{\partial
x^i}-\frac{\partial h_{il}}{\partial x^p}\right)\, . \ea
With the help of the following relations (which are correct up to the order $1/c^4$):
\be{37} \frac{1}{2}\eta^{ij}\eta^{lp}h_{jp}\frac{\partial^2h_{00}}{\partial x^i\partial x^l}\approx\frac{1}{2}h_{11}\triangle
h_{00}=\frac{1}{D-2}\cdot\frac{2}{c^4}\varphi\triangle\varphi\, , \ee
\be{38} \fl\frac{1}{4}\eta^{il}\eta^{np}\left(\frac{\partial h_{n0}}{\partial x^l}-\frac{\partial h_{0l}}{\partial x^n}\right)\left(\frac{\partial h_{p0}}{\partial
x^i}-\frac{\partial h_{i0}}{\partial x^p}\right)\approx\frac{1}{2}\eta^{il}\frac{\partial h_{00}}{\partial x^l}\frac{\partial h_{00}}{\partial
x^i}\approx-\frac{2}{c^4}(\nabla\varphi)^2\, \ee
and
\be{39} \fl\eta^{il}\eta^{np}\frac{\partial h_{00}}{\partial x^n}\left(\frac{\partial h_{pi}}{\partial x^l}+\frac{\partial h_{pl}}{\partial x^i}-\frac{\partial
h_{il}}{\partial x^p}\right)\approx \frac{\partial h_{00}}{\partial x^{\beta}}\frac{\partial}{\partial
x^{\alpha}}\left(2h^{\alpha\beta}-\eta^{\alpha\beta}h_l^l\right)\approx 0\, , \ee
where the condition \rf{14} was used in the latter expression, the 00-component \rf{36} of the Ricci tensor takes the form
\be{40} R_{00}\approx\frac{1}{c^2}\triangle\varphi+\frac{1}{2}\triangle f_{00}+ \frac{1}{D-2}\cdot\frac{2}{c^4}\varphi\triangle\varphi-\frac{2}{c^4}(\nabla\varphi)^2\, .
\ee
The $0\alpha$ component of the Ricci tensor \rf{29} up to the order $1/c^3$ reads
\be{41} \fl R_{0\alpha}\approx-\frac{1}{2}\eta^{il}\frac{\partial^2h_{0\alpha}}{\partial x^i\partial
x^l}+\frac{1}{2}H_{0\alpha}-\frac{1}{2}\eta^{il}\frac{\partial^2f_{0\alpha}}{\partial x^i\partial x^l}+\frac{1}{2}F_{0\alpha}\approx \frac{1}{2}\triangle
f_{0\alpha}+\frac{1}{2c^3}\frac{\partial^2\varphi}{\partial t\partial x^{\alpha}}\, , \ee
where we used formulas \rf{33} and \rf{34}.

Now, we come back to Einstein equation \rf{3}. Substituting \rf{25} and \rf{40} into \rf{3} and taking into account \rf{5} and \rf{16}, we get the following equation for
$f_{00}$:
\ba{42} & &\triangle f_{00}+\frac{1}{D-2}\, \frac{4}{c^4}\varphi\triangle\varphi-\frac{4}{c^4}(\nabla\varphi)^2\nn \\
&=&\frac{2S_DG_{\mathcal{D}}}{c^4}\sum\limits_{p=1}^Nm_p\left(\frac{3D-4}{D-2}\varphi_p+\frac{D}{2(D-2)}v_p^2\right)\delta({\bf r}-{\bf r}_p)\, . \ea
With the help of the auxiliary equation:
\be{43} 4(\nabla\varphi)^2=2\triangle(\varphi^2)-4\varphi\triangle\varphi \ee
and equations \rf{5} and \rf{16}, equation \rf{42} takes the form:
\be{44} \fl\triangle\left(f_{00}-\frac{2}{c^4}\varphi^2\right)=\frac{2S_DG_{\mathcal{D}}}{c^4}\sum\limits_{p=1}^Nm_p
\left(\varphi'_p+\frac{D}{2(D-2)}v_p^2\right)\delta({\bf r}-{\bf r}_p)\, . \ee
Here, $\varphi'_p$ is the potential of the gravitational field in a point with radius vector ${\bf r}_p$ produced by all particles, except of the $ p $-th. Substraction
of the infinite contribution of the gravitational field of the $p$-th particle corresponds to a renormalization of its mass (see \cite{Landau}).
The solution of \rf{44} is:
\be{45} \fl f_{00}=\frac{2}{c^4}\varphi^2({\bf r})+\frac{2}{c^4}\sum\limits_{p=1}^N\varphi'_p\varphi'({\bf r}-{\bf
r}_p)+\frac{D}{D-2}\cdot\frac{1}{c^4}\sum\limits_{p=1}^Nv_p^2\varphi'({\bf r}-{\bf r}_p)\, , \ee
where $\varphi'({\bf r}-{\bf r}_p)$ is the potential of the gravitational field of the $p$-th particle which satisfies the Poisson equation:
\be{46} \triangle\varphi'=\delta^{\alpha\beta}\frac{\partial^2\varphi'}{\partial x^{\alpha}\partial x^{\beta}}=S_DG_{\mathcal{D}}m_p\delta({\bf r}-{\bf r}_p)\, . \ee
It can be easily verified with the help of \rf{5} and \rf{16} that $\varphi'({\bf r}-{\bf r}_p)$ satisfies the condition:
\be{47} \varphi({\bf r})=\sum\limits_{p=1}^N\varphi'({\bf r}-{\bf r}_p)\, . \ee

Therefore, substituting $h_{00} = 2\varphi /c^2$ and $f_{00}$  into \rf{27}, we obtain $g_{00}$ up to the order $1/c^4$:
\ba{48} g_{00}&\approx&1+\frac{2}{c^2}\varphi({\bf r})+\frac{2}{c^4}\varphi^2({\bf r})\nn \\
&+&\frac{2}{c^4}\sum\limits_{p=1}^N\varphi'_p\varphi'({\bf r}-{\bf r}_p)+\frac{D}{D-2}\cdot\frac{1}{c^4}\sum\limits_{p=1}^Nv_p^2\varphi'({\bf r}-{\bf r}_p)\, . \ea
We should mention that the radius vectors ${\bf r}_p$ of the moving gravitating masses depend on time. In this case, potential $\varphi({\bf r})$ in \rf{47} also depends
on time.

The equation for $f_{0\alpha}$ can be obtained by substitution of \rf{26} and \rf{41} into Einstein equation \rf{3}:
\be{49} \triangle f_{0\alpha}=-\frac{2(D-1)}{D-2}\frac{S_DG_{\mathcal{D}}}{c^3}\sum\limits_{p=1}^Nm_pv_{p\alpha}\delta({\bf r}-{\bf
r}_p)-\frac{1}{c^3}\frac{\partial^2\varphi}{\partial t\partial x^{\alpha}}, \ee
whose solution is:
\be{50} f_{0\alpha}=-\frac{2(D-1)}{D-2}\cdot\frac{1}{c^3}\sum\limits_{p=1}^Nv_{p\alpha}\varphi'({\bf r}-{\bf r}_p)-\frac{1}{c^3}\frac{\partial^2f}{\partial t\partial
x^{\alpha}}\, , \ee
where the function $f$ satisfies equation
\be{51} \triangle f=\delta^{\alpha\beta}\frac{\partial^2f}{\partial x^{\alpha}\partial x^{\beta}}=\varphi({\bf r})\, . \ee
Therefore, substituting $h_{0\alpha} = 0$ and $f_{0\alpha}$ into \rf{27}, we get $g_{0\alpha}$ up to the order $1/c^3$:
\be{52} g_{0\alpha}\approx-\frac{2(D-1)}{D-2}\, \frac{1}{c^3}\sum\limits_{p=1}^Nv_{p\alpha}\varphi'({\bf r}-{\bf r}_p)-\frac{1}{c^3}\frac{\partial^2f}{\partial t\partial
x^{\alpha}}\, . \ee

It is necessary to note that in the three-dimensional case $D=3$ \rf{48} and \rf{52} exactly coincide with (106.13) and (106.14) in \cite{Landau} if we take into account
that $\varphi'({\bf r}-{\bf r}_p) = -G_Nm_p/|{\bf r}-{\bf r}_p|$.

From now on we shall consider the case of one gravitating particle of mass $m_1\equiv m$ at rest in our 3-D space but, for generality, moving with constant speed in
extra dimensions. That is $p=1 \Rightarrow \varphi'_1 = 0$ and $v^{\alpha}=dx^{\alpha}/dt = (0,0,0,v_4,v_5,...,v_D)$, where $v_4,v_5,...,v_D$ are constants. In this case
\rf{48} and \rf{52} are reduced correspondingly to
\be{53} g_{00}\approx1+\frac{2}{c^2}\varphi({\bf r})+\frac{2}{c^4}\varphi^2({\bf r})+\frac{Dv^2}{(D-2)c^4}\varphi({\bf r}) \ee
and
\be{53a} g_{0\alpha}\approx-\frac{2(D-1)v_{\alpha}}{(D-2)c^3}\varphi({\bf r})-\frac{1}{c^3}\frac{\partial^2f}{\partial t\partial x^{\alpha}}\, , \ee
where  $\varphi({\bf r}) $ satisfies the Poisson equation:
\be{54} \triangle\varphi=\delta^{\alpha\beta}\frac{\partial^2\varphi}{\partial x^{\alpha}\partial x^{\beta}}=S_DG_{\mathcal{D}}m\delta({\bf r})\, \ee
and $v^2=-g_{\alpha\beta}v^{\alpha}v^{\beta}= v_4^2+v_5^2+...+v_D^2 +O(1/c^2)$ (at the same accuracy $v_{\beta}=-v^{\beta}$). Obviously, the transition to the case where
the gravitating mass is at rest both in our three-dimensional space and in the extra dimensions corresponds to the limit $v_{\alpha} =0 \Rightarrow v^2 =0$. In this case
the potential $\varphi ({\bf r})$ as well as the function $f$ do not depend on time $t$.  We remind that the covariant components $g_{\alpha\beta}$ read (see \rf{17}):
\be{55} g_{\alpha\beta}\approx -\left(1-\frac{1}{D-2}\cdot\frac{2}{c^2}\varphi({\bf r})\right)\delta_{\alpha\beta}\, . \ee

To get all above results, we did not use any concrete form of topology. The only things we used were assumptions of the flatness of metric in the absence of the
gravitating masses and the weakness of the gravitational field and velocities of gravitating masses which perturb the flat metric. Now, to solve \rf{54} we should
specify the topology of space and the boundary conditions. We suppose that the $(D=3+d)$-dimensional space has the factorizable geometry of a product manifold
$M_D=\mathbb{R}^3\times T^{d}$. $\mathbb{R}^3$ describes the three-dimensional flat external (our) space and $T^{d}$ is a torus which corresponds to a $d$-dimensional
internal space with volume $V_d$. For this topology, and with the boundary condition that at infinitely large distances from the gravitating body the potential must go
to the Newtonian expression, we can find the exact solution of the Poisson equation \rf{54} \cite{EZ1,EZ2}. The boundary condition requires that the multidimensional
$G_{\mathcal{D}}$ and Newtonian $G_N$ gravitational constants are connected by the following condition: $S_D G_{\mathcal{D}}/V_d =4\pi G_N$. Assuming that we consider
the gravitational field of a gravitating mass $m$ at distances much greater than periods of tori, we can restrict ourselves to the zero Kaluza-Klein mode. For example,
this approximation is very well satisfied for the planets of the solar system because the inverse-square law experiments show that the extra dimensions in Kaluza-Klein
models should not exceed submillimeter scales \cite{new} (see however \cite{EZ1,EZ2} for models with smeared extra dimensions where Newton's law preserves its shape for
arbitrary distances). Then, the gravitational potential reads
\be{56} \varphi({\bf r})\approx -\frac{G_N m}{r_3} = -\frac{r_gc^2}{2r_3}\, , \ee
where $r_3$ is the length of a radius vector in three-dimensional space and we introduce three-dimensional Schwarzschild radius $r_g=2G_Nm/c^2$. As we mentioned above,
the gravitating mass $m$ is at rest in our three-dimensional space but may move in the extra dimensions. In this case, the extra dimensional components of D-dimensional
radius vector of the gravitating particle depend on time. The exact formulas for the non-relativistic gravitational potential (see \cite{EZ1,EZ2}) show that this
dependence "nests" only in non-zero Kaluza-Klein modes which are exponentially suppressed in considered approximation. Therefore, in this approximation potential
$\varphi({\bf r})$ in \rf{56} does not depend on time.

It is worth noting that all the previous analysis works also in the case when the gravitating masses are uniformly smeared over some or all extra dimensions. Let us take
for simplicity one $(p=1)$ gravitating mass $m_1\equiv m$ which is smeared over all extra dimensions. Obviously, this mass can move only in our usual three dimensions:
$v_1^{\alpha} = dx^{\alpha}_1 /dt =(v_1^1,v_1^2,v_1^3,0, \ldots ,0)$ and its rest mass density \rf{5} now reads:
\be{56a} \rho=\left(m/\prod_{\alpha=1}^d a_{\alpha}\right)\, \delta({\bf r}_3-{\bf r}_{(1)3})\, , \ee
where $a_{\alpha}$ are periods of tori. Then, the solution of the Poisson equation \rf{16} exactly coincides with the Newton potential  if the multidimensional
$G_{\mathcal{D}}$ and Newtonian $G_N$ gravitational constants are connected as $S_D G_{\mathcal{D}}/\prod_{\alpha=1}^d a_{\alpha} =4\pi G_N$ \cite{EZ1,EZ2}. Therefore,
in this case the approximate formula \rf{56} becomes the exact equality:
\be{56b} \varphi({\bf r})=\varphi({\bf r}_3)= -\frac{G_N m}{r_3} = -\frac{r_gc^2}{2r_3}\, . \ee

In the approximation \rf{56} (or with \rf{56b} for "smeared" extra dimensions), the covariant components \rf{53}, \rf{53a} and \rf{55} take the form
\ba{57}
g_{00}&\approx & 1-\frac{r_g}{r_3}+\frac{r_g^2}{2r_3^2}-\frac{Dv^2}{2(D-2)c^2}\frac{r_g}{r_3}\, , \nn\\
g_{0\alpha}&\approx & \frac{(D-1)v_{\alpha}}{(D-2)c}\frac{r_g}{r_3},\ \ \ g_{\alpha\beta}\approx-\left(1+\frac{1}{D-2}\cdot\frac{r_g}{r_3}\right)\delta_{\alpha\beta}\, .
\ea
For the contravariant components we obtain:
\ba{58}
g^{00}&\approx&1+\frac{r_g}{r_3}+\frac{r_g^2}{2r_3^2}+\frac{Dv^2}{2(D-2)c^2}\frac{r_g}{r_3}\, ,\nn \\
g^{0\alpha}&\approx&-\frac{(D-1)v^{\alpha}}{(D-2)c}\frac{r_g}{r_3},\ \ \ g^{\alpha\beta}\approx-\left(1-\frac{1}{D-2}\cdot\frac{r_g}{r_3}\right)\delta_{\alpha\beta}\, .
\ea
It is not difficult to verify that these components satisfy the condition:
\be{58a} g_{ik}g^{kj} = \left(\begin{array}{ccc}
1+O(1/c^6)& 0+O(1/c^5)\\
0+O(1/c^5)& \delta_{\alpha\beta} + O(1/c^4)
\end{array}\right)\, .
\ee

The metric components \rf{57} demonstrate that in this approximation the spacial section $t=const$ is conformal to the Euclidean metric. Hence, the spacial coordinates
are isotropic ones. It is convenient to use three-dimensional spherical coordinates $r_3,\theta,\psi $ instead of the Cartesian coordinates $x^1\equiv x,x^2\equiv
y,x^3\equiv z$. In these coordinates
the metric \rf{1} reads:
\ba{59} ds^2&\approx&\left(1-\frac{r_g}{r_3}+\frac{r_g^2}{2r_3^2}-\frac{Dv^2}{2(D-2)c^2}\frac{r_g}{r_3} \right)c^2dt^2\nn \\
&+&\frac{2(D-1)}{(D-2)c}\frac{r_g}{r_3}c dt\sum_{\alpha=4}^Dv_{\alpha}dx^{\alpha}\nn \\
&-&\left(1+\frac{1}{D-2}\, \frac{r_g}{r_3}\right)\left(dr_3^2+r_3^2d\theta^2+r_3^2\sin^2\theta d\psi^2\right)\nn \\
&-&\left(1+\frac{1}{D-2}\, \frac{r_g}{r_3}\right)\left((dx^4)^2+(dx^5)^2+\ldots + (dx^D)^2\right)\, . \ea
As we mentioned above, this metric corresponds to a gravitating mass in the rest in our three dimensional space. If the mass is smeared over extra dimensions, the
appropriate velocity components vanish.


\section{\label{sec:3}Classical gravitational tests}

\setcounter{equation}{0}

Now, we want to check the obtained above  multidimensional metric \rf{59} from the point of its consistency with the famous classical tests: frequency shift, perihelion
shift, deflection of light and time delay of radar echoes (the Shapiro time delay effect). We also want to calculate the parameterized post-Newtonian (PPN) parameters for obtained metric coefficients. It is well known that four-dimensional general relativity is in good agreement with these experiments and observed PPN parameters. Can the considered Kaluza-Klein models with point-like sources also be in concordance with observations?

\subsection{Frequency shift}

To investigate the gravitational redshift formula in the spacetime \rf{59}, we can use the famous expression for relation between the frequency  $\omega_1$ of a light
signal emitted at a point 1 with the metric component $\left.g_{00}\right|_1$ and the frequency $\omega_2$ received at a point 2 with the metric component
$\left.g_{00}\right|_2$:
\be{f1} \omega_1\left[\left(g_{00}\right)^{1/2}\right]_1=\omega_2\left[\left(g_{00}\right)^{1/2}\right]_2\, . \ee
Therefore, up to the order $1/c^2$ we get
\be{f2} \omega_2\approx\omega_1\left(1+\frac{\varphi_1-\varphi_2}{c^2}\right)\, , \ee
where non-relativistic potential $\varphi$ is given by \rf{56}. In considered approximation, this formula exactly coincides with the one from general relativity.
Therefore, for this type of experiments it is hardly possible to observe the difference between the usual four-dimensional general relativity and multidimensional
Kaluza-Klein models.

\subsection{Perihelion shift}

Let us consider now the motion of a test body of mass $m'$ in the gravitational field described by metric \rf{59}. The Hamilton-Jacobi equation
\be{60} g^{ik}\frac{\partial S}{\partial x^i}\frac{\partial S}{\partial x^k}-m'^2c^2=0 \ee
for this test body moving in the orbital plane $\theta=\pi/2$ reads
\ba{61} \fl\frac{1}{c^2}\left(1+\frac{r_g}{r_3}+\frac{r_g^2}{2r_3^2}+\frac{Dv^2}{2(D-2)c^2}\frac{r_g}{r_3}\right)\left(\frac{\partial S}{\partial t}\right)^2
-\frac{2(D-1)v^{\alpha}}{(D-2)c^2}\frac{r_g}{r_3}\, \frac{\partial S}{\partial t}\frac{\partial
S}{\partial x^{\alpha}}\nn \\
\fl-\left(1-\frac{1}{D-2}\, \frac{r_g}{r_3}\right)\left(\frac{\partial S}{\partial r_3}\right)^2
-\frac{1}{r_3^2}\left(1-\frac{1}{D-2}\, \frac{r_g}{r_3}\right)\left(\frac{\partial S}{\partial \psi}\right)^2\nn \\
\fl-\left(1-\frac{1}{D-2}\, \frac{r_g}{r_3}\right)\left[\left(\frac{\partial S}{\partial x^4}\right)^2+...+\left(\frac{\partial S}{\partial x^D}\right)^2\right] -
m'^2c^2\approx 0\, . \ea
We investigate this equation by separation of variables considering the action in the form
\be{61a} S=-E't+M\psi+S_{r_3}(r_3)+S_4\left(x^4\right)+...+S_D\left(x^D\right)\, . \ee
Here, $E'\approx m'c^2+E$ is the energy of the test body, which includes the rest energy $m'c^2$ and non-relativistic energy $E$, and $M$ is the angular momentum.
Substituting this expression for the action $S$ in the formula \rf{61}, we obtain an expression for $(dS_{r_3}/dr_3)^2$  holding there the terms up to the order $1/c^2$:
\ba{62}
\fl\left(\frac{dS_{r_3}}{dr_3}\right)^2 \approx\frac{E'^2}{c^2}\left(1-\frac{1}{D-2}\, \frac{r_g}{r_3}\right)^{-1}\left(1+\frac{r_g}{r_3}+\frac{r_g^2}{2r_3^2} +\frac{Dv^2}{2(D-2)c^2}\frac{r_g}{r_3}\right)\nn \\
\fl-\frac{M^2}{r_3^2}
+E'\left(1-\frac{1}{D-2}\, \frac{r_g}{r_3}\right)^{-1}\frac{2(D-1)v^{\alpha}}{(D-2)c^2}\frac{r_g}{r_3}\, \frac{\partial S}{\partial x^{\alpha}}\nn \\
\fl-\left(\frac{dS_4}{dx^4}\right)^2-\ldots -\left(\frac{dS_D}{dx^D}\right)^2 -m'^2c^2\left(1-\frac{1}{D-2}\, \frac{r_g}{r_3}\right)^{-1}\nn \\
\fl\approx \left(2m'E-\left(p_4^2+...+p_D^2\right)+\frac{E^2}{c^2}\right)-\frac{1}{r_3^2}\left(M^2-\frac{Dm'^2c^2r_g^2}{2(D-2)}\right)\\
\fl+\frac{1}{r_3}\left(m'^2c^2r_g+\frac{2(D-1)}{D-2}\, m'Er_g+\frac{D}{2(D-2)}\, m'^2r_gv^2+\frac{2(D-1)}{(D-2)}\, m'r_g\sum_{\alpha=4}^Dv^{\alpha}p_{\alpha}\right)\nn
\, , \ea
where $p_{\alpha} = \partial S/\partial x^{\alpha}=dS_{\alpha}/dx^{\alpha}\; (\alpha = 4,\ldots ,D)$ are the components of momentum of the test body in the extra
dimensions. If the gravitating and test masses are localized on the same brane then these components are equal to zero. Integrating the square root of this expression
with respect to $r_3$, we get $S_{r_3}$ in the following form:
\ba{63}
\fl S_{r_3}\approx\int \left[\left(2m'E-\left(p_4^2+...+p_D^2\right)+\frac{E^2}{c^2}\right)\right.\nn \\
\fl+\frac{1}{r_3}\left(m'^2c^2r_g+\frac{2(D-1)}{D-2}\, m'Er_g+\frac{D}{2(D-2)}\, m'^2r_gv^2+\frac{2(D-1)}{(D-2)}\,
m'r_g\sum_{\alpha=4}^Dv^{\alpha}p_{\alpha}\right)\nn \\
\fl-\left.\frac{1}{r_3^2}\left(M^2-\frac{Dm'^2c^2r_g^2}{2(D-2)}\right) \right]^{1/2}dr_3\, , \ea

It is well known (see e.g. $\S$ 47 in \cite{Landau mechanics}) that for any integral of motion $I$ of a system with action $S$ the following equation should hold:
\be{64a} \frac{\partial S}{\partial I}=const\, . \ee
Because the angular momentum $M$ is the integral of motion, the trajectory of the test body is defined by the equation
\be{64} \frac{\partial S}{\partial M}=\psi+\frac{\partial S_{r_3}}{\partial M}=\mathrm{const}\, , \ee
where we use \rf{61a}.

Let now the Sun be the gravitating mass and the planets of the solar system be the test bodies. Then, the change of the angle during one revolution of a planet on an
orbit is
\be{65} \Delta\psi=-\frac{\partial}{\partial M}\Delta S_{r_3}\, , \ee
where $\Delta S_{r_3}$ is the corresponding change of $S_{r_3}$. It is well known that the perihelion shift originates due to small relativistic correction $\varepsilon$
to $M^2$ in $S_{r_3}$: $\quad M^2/r^2_3 \Rightarrow (M^2 -\varepsilon)/r^2_3$. \rf{63} shows that in our case $\varepsilon = Dm'^2c^2r_g^2/[2(D-2)]$. Expanding $S_{r_3}$
in powers of this correction:
\ba{66} S_{r_3}&=&S_{r_3}(M^2-\varepsilon) \approx S_{r_3}^{(0)}-\varepsilon \frac{\partial S_{r_3}^{(0)}}{\partial M^2}\nn \\
&=& S_{r_3}^{(0)}-\frac{\varepsilon}{2M} \frac{\partial S_{r_3}^{(0)}}{\partial M}=S_{r_3}^{(0)}-\frac{Dm'^2c^2r_g^2}{4(D-2)M}\frac{\partial S_{r_3}^{(0)}}{\partial M}\,
, \ea
 where $S_{r_3}^{(0)}\equiv S_{r_3}(M^2)$, we obtain
\be{67} \Delta S_{r_3}\approx\Delta S_{r_3}^{(0)}-\frac{D m'^2c^2r_g^2}{4(D-2)M}\frac{\partial\Delta S_{r_3}^{(0)}}{\partial M}\, . \ee
Differentiating this equation with respect to $M$ we get
\be{68} \Delta\psi\approx2\pi+\frac{D\pi m'^2c^2r_g^2}{2(D-2)M^2}\, , \ee
where we took into account $-\partial \Delta S_{r_3}^{(0)}/\partial M=\Delta\psi^{(0)}=2\pi$. Therefore, the second term in \rf{68} gives the required formula for the
perihelion shift in our multidimensional case:
\be{69} \delta\psi =\frac{D\pi m'^2c^2r_g^2}{2(D-2)M^2}=\frac{D\pi r_g}{(D-2)a(1-e^2)}\, , \ee
where in this equation we used the well-known relation $M^2=m'^{2}r_{g}c^2a(1-e^2)/2$ with $a$ and $e$ being the semi-major axis and the eccentricity of the ellipse,
respectively. For the three-dimensional case $D=3$, this equation exactly coincides with formula (101.7) in \cite{Landau}. It can be easily seen that the result \rf{69}
does not depend on motion of the gravitating and test masses in the extra dimensions.

It make sense to apply this formula to Mercury because in the solar system it has the most significant discrepancy between the measurement value of the perihelion shift
and its calculated value using Newton's formalism. The observed discrepancy is $43.11\pm 0.21$ arcsec per century. This missing value is usually explained by the
relativistic effects of the form of \rf{69}.
However, only in three-dimensional case $D=3$ \rf{69} gives the satisfactory result $42.94''$ which is within the measurement accuracy. For $D=4$ and $D=9$ models we
obtain $28.63''$ and $18.40''$, respectively, which are very far from the observable value.

\subsection{Deflection of light}

Let us consider now the propagation of light in gravitational field with metric \rf{59}. In the case of massless particles, the Hamilton-Jacobi equation \rf{60} is
reduced to the eikonal equation:
\be{70} g^{ik}\frac{\partial\Psi}{\partial x^i}\frac{\partial\Psi}{\partial x^k}=0\, , \ee
which for the metric \rf{59} reads
\ba{71} \fl\frac{1}{c^2}\left(1+\frac{r_g}{r_3}+\frac{r_g^2}{2r_3^2}+\frac{Dv^2}{2(D-2)c^2}\frac{r_g}{r_3}\right)\left(\frac{\partial\Psi}{\partial t}\right)^2
-\frac{2(D-1)v^{\alpha}}{(D-2)c^2}\frac{r_g}{r_3}\, \frac{\partial\Psi}{\partial
t}\frac{\partial\Psi}{\partial x^{\alpha}}\nn \\
\fl-\left(1-\frac{1}{D-2}\, \frac{r_g}{r_3}\right)\left(\frac{\partial\Psi}{\partial
r_3}\right)^2-\frac{1}{r_3^2}\left(1-\frac{1}{D-2}\, \frac{r_g}{r_3}\right)\left(\frac{\partial\Psi}{\partial \psi}\right)^2\nn \\
\fl-\left(1-\frac{1}{D-2}\, \frac{r_g}{r_3}\right)\left[\left(\frac{\partial\Psi}{\partial x^4}\right)^2+...+\left(\frac{\partial\Psi}{\partial
x^D}\right)^2\right]\approx 0\, , \ea
where we take into account that light propagates in the orbital plane $\theta =\pi/2$. The eikonal function $\Psi$ can be written in the form
\be{72} \fl\Psi=-\omega_0t+\frac{\rho\omega_0}{c}\psi+\Psi_{r_3}(r_3)+\Psi_4\left(x^4\right)+\Psi_5\left(x^5\right)+...+\Psi_D\left(x^D\right)\, , \ee
where $\omega_0 = -\partial\Psi/\partial t$ is the frequency of light and $\rho$ is a constant. Later we will show that $\rho$ is the impact parameter, i.e. distance of
closest approach of the ray's path to the gravitating mass. Taking into account that $k=\omega_0/c$ is the absolute value of the wave-vector, it is clear that $M \equiv
\rho k = \rho\omega_0/c$ plays the role of the angular momentum for the light beam.

Now we consider the natural case when the light propagates in our three-dimensional space and does not have components of momentum in the extra dimensions, that is
$p_{\alpha}=d\Psi_{\alpha}/dx^{\alpha}\equiv 0\, , \quad \alpha =4,\ldots ,D$. Then from \rf{71}, using \rf{72}, we obtain up to the order $O(1/c^4)$ the following
formula:
\ba{73} \fl\left(\frac{d\Psi_{r_3}}{dr_3}\right)^2\approx\frac{\omega_0^2}{c^2}\left(1-\frac{1}{D-2}
\, \frac{r_g}{r_3}\right)^{-1}\left(1+\frac{r_g}{r_3}+\frac{r_g^2}{2r_3^2}+\frac{Dv^2}{2(D-2)c^2}\frac{r_g}{r_3}\right)-\frac{\rho^2\omega_0^2}{c^2r_3^2}\nn \\
\fl\approx\frac{\omega_0^2}{c^2}\left(1+\frac{D-1}{D-2}\, \frac{r_g}{r_3}-\frac{\rho^2}{r_3^2}\right)\, . \ea
Integrating this expression we get:
\be{74} \Psi_{r_3}\approx\frac{\omega_0}{c}\int\left(1+\frac{D-1}{D-2}\, \frac{r_g}{r_3}-\frac{\rho^2}{r_3^2}\right)^{1/2}dr_3\, . \ee
Considering the term with $r_g/r_3$ as a small relativistic correction, we expand the integrand up to the order $O(1/c^3)$:
\ba{75} \Psi_{r_3}&\approx&\Psi_{r_3}^{(0)}+\frac{D-1}{2(D-2)}\, \frac{r_g\omega_0}{c}\int\left(r_3^2-\rho^2\right)^{-1/2}dr_3\nn \\
&=&\Psi_{r_3}^{(0)}+\frac{D-1}{2(D-2)}\, \frac{r_g\omega_0}{c}\mathrm{arccosh}\frac{r_3}{\rho}\, , \ea
where the non-relativistic (i.e. gravity is absent: $r_g\equiv 0$) eikonal function is
\be{76} \Psi_{r_3}^{(0)}=\frac{\omega_0}{c}\int\left(1-\frac{\rho^2}{r_3^2}\right)^{1/2}dr_3 \equiv
\int\left(\left(\frac{\omega_0}{c}\right)^2-\frac{M^2}{r_3^2}\right)^{1/2}dr_3\, . \ee
For this non-relativistic approximation the trajectory of the light beam is a straight line. Indeed, in this case (by full analogy with \rf{64}) we have
\be{77} \frac{\partial \Psi^{(0)}}{\partial M}=\psi^{(0)}+\frac{\partial \Psi^{(0)}_{r_3}}{\partial M}=\psi^{(0)} - \arccos (\rho/r_3)=0\, , \ee
where the constant is taken in such a way that $\psi^{(0)} \to \pi/2$ for $r_3\to \infty$. Thus, the trajectory $\rho =r_3 \cos \psi^{(0)}$ is the straight line.
Obviously, in the non-relativistic case the total change of the angle $\psi^{(0)} $ is $\Delta\psi^{(0)}=-\partial\Delta\Psi_{r_3}^{(0)}/\partial M=\pi$.

Coming back to the relativistic case \rf{75}, for the light beam travelling from some distance $r_3=R$ to the closest approach to the gravitating mass  at $r_3=\rho$ and
again to the distance $r_3=R$, the change of the eikonal function is
\be{78} \Delta\Psi_{r_3}\approx\Delta\Psi_{r_3}^{(0)}+\frac{D-1}{D-2}\, \frac{r_g\omega_0}{c}\mathrm{arccosh}\frac{R}{\rho}\, . \ee
The corresponding change of the polar angle $\psi$ is
\ba{79}
\frac{\partial \Psi}{\partial M}=\psi+\frac{\partial \Psi_{r_3}}{\partial M}=\mathrm{const} \nn \\
\fl\Longrightarrow \quad \Delta\psi=-\frac{\partial\Delta\Psi_{r_3}}{\partial M}\approx-\frac{\partial\Delta\Psi_{r_3}^{(0)}}{\partial M}+\frac{D-1}{D-2}\,
\frac{r_gR}{\rho}\left(R^2-\rho^2\right)^{-1/2}\, . \ea
Thus in the limit $R\rightarrow+\infty$ we finally get:
\be{80} \Delta\psi\approx\pi+\frac{D-1}{D-2}\, \frac{r_g}{\rho}\, . \ee
Therefore, the second term in \rf{80} gives the required formula for the deflection of light in our multidimensional case:
\be{81} \delta \psi = \frac{D-1}{D-2}\, \frac{r_g}{\rho} \ee
For the three-dimensional case $D=3$, this equation exactly coincides with formula (101.9) in \cite{Landau}.

Now we apply this formula to the Sun. Obviously, the radius $R_{Sun}$ of the Sun is much greater than the size of the extra dimensions and approximation \rf{56} works well on
the distances $r_3\ge R_{Sun}$. For general relativity and for a ray that grazes the Sun's limb (i.e. $\rho \approx R_{Sun}$) $\delta \psi \approx 1.75$ arcsec which is in very good agreement with
observational data \cite{light deflection}. \rf{81} shows that we get this value of $\delta \psi$  only for usual three-dimensional space. In the case $D=4$ and $D=9$ we
obtain correspondingly $\delta \psi\approx 1.31^{''}$ and $\delta \psi\approx 1.00^{''}$ which are very far from the observable value.

\subsection{Parameterized post-Newtonian Parameters and gravitational tests}

It is well known (see e.g. \cite{Will,Straumann}) that in PPN formalism the static, spherically symmetric metric in isotropic
coordinates reads
\be{p.1} ds^2=\left(1-\frac{r_g}{r_3}+\beta\frac{r_g^2}{2r_3^2} \right)c^2dt^2
-\left(1+\gamma \frac{r_g}{r_3}\right)\sum_{i=1}^{3}\left(dx^i\right)^2
\, .
\ee
In general relativity, $\beta=\gamma=1$. However, simple comparison of equations \rf{p.1} and \rf{59} shows that the PPN parameters $\beta$ and $\gamma$ in our case are
\be{p.2}
\beta =1\; , \qquad \gamma =\frac{1}{D-2}\, .
\ee
The latter expression shows that parameter $\gamma$ coincides with the corresponding value in general relativity if $D=3$. Only in this case $\gamma =1$. According to
the experimental data, $\gamma $ should be very close to 1. The tightest constraint on $\gamma$ comes from  the Shapiro time-delay experiment using the Cassini
spacecraft: $\gamma-1 =(2.1\pm 2.3)\times 10^{-5}$ \cite{Bertotti,Will2,JKh}. On the other hand, for $D=4,9$ we get from \rf{p.2} that $\gamma-1=-1/2,-6/7$ respectively,
which is very far from the experimental data.

The formulas of the gravitational tests can be expressed via the PPN parameters \cite{Will,Will2}. For example, the perihelion shift and the deflection of light read
correspondingly:
\ba{p.3}
\delta \psi &=& \frac13 \left(2+2\gamma -\beta\right)\, \frac{3\pi r_g}{a(1-e^2)}\, , \\
\delta \psi &=& \left(1+\gamma\right)\, \frac{r_g}{\rho}\, .
\ea
Now, if we substitute in these expressions the values from equation \rf{p.2}  for $\beta $ and $\gamma$, then we exactly restore our formulas \rf{69}
and \rf{81}.

It makes sense to present the expression for the \emph{time delay of radar echoes} (the Shapiro time delay effect) via the PPN parameters. This effect consists in time
difference of propagation of electromagnetic signals between two points (or for a round trip) in the curved and flat spaces. Usually, a signal transmits from the Earth
through a region near the Sun to another planet or satellite and then reflects back to the Earth. If the planet (or satellite) is on the far side of the Sun from the
Earth (superior conjunction), then the formula for the time delay reads \cite{Will,Will2}:
\be{p.5}
\delta t = (1+\gamma)\frac{r_g}{c}\ln\left(\frac{4r_{Earth}\, r_{planet}}{R_{Sun}^2}\right)\, ,
\ee
where $r_{Earth}$ and $r_{planet}$ are the distances from the Sun to the Earth and to the planet, respectively.
If we put into this formula the value of $\gamma$ from equation \rf{p.2}, we get
\be{p.6}
\delta t = \frac{D-1}{D-2}\frac{r_g}{c}\ln\left(\frac{4r_{Earth}\, r_{planet}}{R_{Sun}^2}\right)\, .
\ee
Obviously, this formula coincides with the general relativity  only for $D=3$ (see e.g. \cite{Straumann}). For all others
values of $D$ the time delay differs from the general relativity by the factor of $O(1)$.


\section{Conclusion}

In our paper we investigated classical gravitational tests (frequency shift, perihelion shift, deflection of light and time delay of radar echoes) for multidimensional
models with compact internal spaces in the form of tori. We supposed that in the absence of gravitating masses the metric is a flat one. Gravitating point-like masses
(moving or at rest) perturb this metric and we considered these perturbations in a weak field approximation. In this approximation, we obtained the asymptotic form of
the metric coefficients. Until this point we did not require the compactness of the extra dimensions. This approach is valid for any number of spatial dimensions $D\ge
3$ and generalizes well-known calculations \cite{Landau} in four-dimensional space-time. Then, we admitted that, first, the extra dimensions are compact and have the
topology of tori and, second, gravitational potential far away from gravitating masses tends to non-relativistic Newtonian limit. It gave us a possibility to specify the
non-relativistic gravitational potential for considered models. In turn, it enabled us to specify the metric coefficients. In the case of a gravitating delta-shaped body
at rest, we used these metric coefficients to calculate frequency shift, perihelion shift, deflection of light and parameterized post-Newtonian parameters $\beta$ and
$\gamma$. With the help of PPN parameter $\gamma$, we also obtain the formula for the time delay of radar echoes.  We demonstrated that for the frequency shift type
experiment it is hardly possible to observe the difference between the usual four-dimensional general relativity and multidimensional Kaluza-Klein models. However, the
situation is quite different for perihelion shift, deflection of light and the Shapiro time delay effect. In these three cases we obtained formulas which generalize the
corresponding ones in general relativity. We showed that all of these formulas
depend on a total number of spatial dimensions $D$ and they are in good agreement with observations only in ordinary
three-dimensional space $D=3$. This result does not depend explicitly on the size of the extra dimensions. Therefore, it is impossible to avoid the problem with
classical gravitational tests in a limit of arbitrary small sizes of the extra dimensions.

Therefore, our results show that in considered multidimensional Kaluza-Klein models the point-like gravitating masses cannot produce gravitational field which
corresponds to the classical gravitational tests. Moreover, it is not difficult to show (see our forthcoming paper), that similar problem arises in the case of a compact
static spherically symmetric perfect fluid with the following conditions for the energy-momentum tensor: $T_{00}\gg T_{0\alpha}, T_{\alpha\beta}\, , \quad \alpha,\beta
=1,\ldots ,D$. To avoid this problem, it is necessary to break a symmetry between our three usual spatial dimensions and the extra dimensions. The branes are among the
most natural candidates for solving this problem. Our results work in favor of the brane-world models. It is of interest also to check models with a non-linear action
$f(R)$. However, to prove viability of these models it is necessary to perform the similar investigations.

\ack We want to thank Uwe G\"unther and
Ignatios Antoniadis for useful comments.
This work was supported in part by the "Cosmomicrophysics" programme of the Physics and Astronomy
Division of the National Academy of Sciences of Ukraine.  A. Zh. acknowledges the hospitality
of the Theory Division of CERN and the High Energy, Cosmology and Astroparticle Physics Section of the ICTP
during final preparation of this work.


\end{document}